\documentclass[10pt, conference]{IEEEtran}
\IEEEoverridecommandlockouts
\usepackage{cite}
\usepackage{amsmath,amssymb,amsfonts}
\usepackage{graphicx}
\usepackage{textcomp}
\usepackage{xcolor}
\usepackage{bm}
\usepackage{dsfont}
\usepackage{comment}
\usepackage{algorithm}
\usepackage{algpseudocode}
\usepackage{tikz}
\usepackage[frozencache=true,cachedir=minted-cache]{minted} 
\usepackage{listings}
\usepackage{multirow}

\usepackage{xcolor}
\usepackage{scalerel}
\usepackage{tikz}
\usetikzlibrary{svg.path}

\definecolor{orcidlogocol}{HTML}{A6CE39}
\tikzset{
  orcidlogo/.pic={
    \fill[orcidlogocol] svg{M256,128c0,70.7-57.3,128-128,128C57.3,256,0,198.7,0,128C0,57.3,57.3,0,128,0C198.7,0,256,57.3,256,128z};
    \fill[white] svg{M86.3,186.2H70.9V79.1h15.4v48.4V186.2z}
                 svg{M108.9,79.1h41.6c39.6,0,57,28.3,57,53.6c0,27.5-21.5,53.6-56.8,53.6h-41.8V79.1z M124.3,172.4h24.5c34.9,0,42.9-26.5,42.9-39.7c0-21.5-13.7-39.7-43.7-39.7h-23.7V172.4z}
                 svg{M88.7,56.8c0,5.5-4.5,10.1-10.1,10.1c-5.6,0-10.1-4.6-10.1-10.1c0-5.6,4.5-10.1,10.1-10.1C84.2,46.7,88.7,51.3,88.7,56.8z};
  }
}

\newcommand\orcidicon[1]{\href{https://orcid.org/#1}{\mbox{\scalerel*{
\begin{tikzpicture}[yscale=-1,transform shape]
\pic{orcidlogo};
\end{tikzpicture}
}{|}}}}
\usepackage{hyperref}

\definecolor{LightGray}{gray}{0.9}
\definecolor{darkspringgreen}{rgb}{0.09, 0.45, 0.27}
\def\BibTeX{{\rm B\kern-.05em{\sc i\kern-.025em b}\kern-.08em
    T\kern-.1667em\lower.7ex\hbox{E}\kern-.125emX}}

\makeatletter
\let\@float@c@listing\@caption
\makeatother
\newcommand\copyrighttext{%
  \footnotesize \textcopyright 2023 IEEE. Personal use of this material is permitted.
  Permission from IEEE must be obtained for all other uses, in any current or future
  media, including reprinting/republishing this material for advertising or promotional
  purposes, creating new collective works, for resale or redistribution to servers or
  lists, or reuse of any copyrighted component of this work in other works.
  DOI: 10.1109/DAC56929.2023.10247664.}
\newcommand\copyrightnotice{%
\begin{tikzpicture}[remember picture,overlay]
\node[anchor=south,yshift=10pt] at (current page.south) {\fbox{\parbox{\dimexpr\textwidth-\fboxsep-\fboxrule\relax}{\copyrighttext}}};
\end{tikzpicture}%
}
\begin{document}
\bstctlcite{IEEEexample:BSTcontrol}
\title{HTVM: Efficient Neural Network Deployment On Heterogeneous TinyML Platforms
\footnotesize \thanks{This project has been partly funded by the Fund for Scientific Research Flanders (FWO Vlaanderen), the EU Horizon 2020 programme under grant agreement No. 101070374 and the Flanders AI Research Program (FAIR).}}
\renewcommand{\baselinestretch}{0.933}

\author{
\IEEEauthorblockN{
Josse Van Delm\textsuperscript{1}\orcidicon{0000-0002-9503-403X}, 
Maarten Vandersteegen\textsuperscript{1}\orcidicon{0000-0002-9377-0922},
Alessio Burrello\textsuperscript{2}\orcidicon{0000-0002-6215-8220},
Giuseppe Maria Sarda\textsuperscript{1,3}\orcidicon{0000-0001-6231-3553}, \\
Francesco Conti\textsuperscript{2}\orcidicon{0000-0002-7924-933X},
Daniele Jahier Pagliari\textsuperscript{4}\orcidicon{0000-0002-2872-7071},
Luca Benini\textsuperscript{2}\orcidicon{0000-0001-8068-3806},
Marian Verhelst\textsuperscript{1,3}\orcidicon{0000-0003-3495-9263}
}
\IEEEauthorblockA{
\textsuperscript{1}\textit{KU Leuven}, Belgium \textsuperscript{2}\textit{University of Bologna}, Italy, \textsuperscript{3}\textit{imec}, Belgium, \textsuperscript{4}\textit{Politecnico di Torino}, Italy}{}
}

\maketitle
\copyrightnotice
\begin{abstract}
Optimal deployment of deep neural networks (DNNs) on state-of-the-art Systems-on-Chips (SoCs) is crucial for tiny machine learning (TinyML) at the edge. The complexity of these SoCs makes deployment non-trivial, as they typically contain multiple heterogeneous compute cores with limited, programmer-managed memory to optimize latency and energy efficiency.
We propose HTVM – a compiler that merges TVM with DORY to maximize the utilization of heterogeneous accelerators and minimize data movements. HTVM allows deploying the MLPerf™ Tiny suite on DIANA, an SoC with a RISC-V CPU, and digital and analog compute-in-memory AI accelerators, at 120x improved performance over plain TVM deployment.
\end{abstract}

\section{Introduction}
Nowadays, new smart digital applications increasingly rely on near-sensor data processing to meet privacy, latency, and energy requirements.
To support this edge computing paradigm for deep learning, embedded Systems-on-Chips (SoCs) are enhanced with one or more on-chip hardware accelerators \cite{genc2021gemmini, ueyoshi2022diana, jain2022tinyvers}.
These accelerators can efficiently perform inference of (Deep) Neural Networks ((D)NNs), reducing energy consumption by more than one order of magnitude compared to general-purpose processors \cite{accelerator_comparison}.
However, such on-chip accelerators usually support only a limited set of DNN operators with specific constraints related to (low) bit precision, data layout, or dimensions. Moreover, they typically only have small onboard memories \cite{burrello2021dory}.

Deploying neural networks on such tiny machine learning (TinyML) platforms is a daunting task for application developers since in-depth hardware-specific knowledge is required to reach the SoC's full potential.
This problem exacerbates for new advanced platforms since they entail complex memory schemes and combine different specialized DNN accelerators.

Recently, many automated DNN deployment toolchains have been proposed to cope with this problem, aiding developers to map their DNNs on edge platforms \cite{burrello2021dory, chen2018tvm}.
However, off-the-shelf deployment toolchains are usually limited either in generality - targeting specific SoCs or accelerator architectures - or in performance - unable to exploit dedicated accelerator hardware maximally due to their generality.

In this work, we aim to fill this gap by proposing \textit{HTVM}, a hybrid deployment toolchain. HTVM can efficiently deploy DNNs on modern TinyML platforms consisting of a microcontroller (MCU) CPU core, multiple accelerator cores with varying data flows and sizes, and an N-level memory system. 

Our contributions towards this goal are:
(1) We extend the TVM compilation flow with a memory-planning back-end (DORY \cite{burrello2021dory}) that generates code and optimizes data movement for dedicated accelerator hardware.
The proposed flow works entirely ahead of time and enables efficient hardware acceleration of a wide variety of DNN layers on memory-constrained devices through hardware-aware layer tiling.
(2) We validate end extensively benchmark our approach by deploying various layers on DIANA, a heterogeneous processing platform encapsulating a RISC-V host, a 500k MAC/cycle analog-in-memory-compute accelerator, and a digital DNN accelerator with a 2-level memory system \cite{ueyoshi2022diana}.
Our hardware-aware tiling approach enables the tiled execution of large layers and achieves up to 6.2$\times$ speed-up over hardware-agnostic tiling. 
Also, it achieves respective performance levels on average only 15.52\% / 5.19\% less than the digital/analog accelerator theoretical peak performance for convolutional layers.
(3) With HTVM, we deploy end-to-end networks of the MLPerf\textsuperscript{TM} Tiny suite on DIANA and compare it with other recently published results.
By exploiting coarse-grained accelerator instructions and a low-overhead runtime, the generated code can reduce overall binary size by up to 12.3\% at equal bit precision compared to plain TVM. By combining multiple accelerators, we need to dispatch fewer kernels from the networks to the general-purpose CPU, decreasing the total latency by up to 8$\times$ over digital- or analog-only single-accelerator solutions.
Our code is open-source at \url{https://github.com/KULeuven-MICAS/htvm}.

The rest of the paper is organized as follows.
In Sec. \ref{sec:related}, we discuss emerging hardware architectures and new deployment tools.
Sec. \ref{sec:methods} details our contribution, while Sec. \ref{sec:results} depicts the results of HTVM. In this section, we use the DIANA SoC as a benchmarking platform. However, HTVM is general enough to support a new off-the-shelf heterogeneous platform.
Sec. \ref{sec:conclusion} concludes the paper with final remarks.
\section{Background and Related Works}
\label{sec:related}
Hardware-software co-design is a crucial ingredient for optimizing DNN inference at the edge. 
In this section, we detail new AI-oriented hardware SoCs based on heterogeneous architectures and software deployment stacks that cater for this emerging class of heterogeneous multi-accelerator platforms.

\subsection{Heterogeneous platforms for edge DNN inference}
The breakdown of Dennard scaling has driven hardware designers towards more specialized processor designs for increased system performance. 
For DNN inference, this offers excellent specialization opportunities since highly parallel workloads (like Conv2D and GEMM) are prevalent. 
Recently, to exploit this opportunity, there has been a trend towards heterogeneous computation, i.e., towards platforms that integrate a CPU with different hardware accelerators/cores with different trade-offs in terms of accuracy, latency, or energy consumption.
The goal is to realize platforms that can switch between different computation domains based, for instance, on the target application or the input complexity, for better accuracy, higher energy efficiency, or lower latency. 

There are many examples in both industry~\cite{xavier_dac22} and academia~\cite{hp_ssc2020, ima_ssc2022, ueyoshi2022diana} that embody these concepts.
For instance, the commercially available Jetson AGX Xavier from NVIDIA includes an 8-core ARM CPU, an NVIDIA Volta GPU, and 2 deep learning accelerators (NVDLA).
Running layers on NVDLA or GPU allows running end-to-end networks at lower latency or energy consumption~\cite{xavier_dac22}. 
Notably, while the GPU can execute every workload, the NVDLA can not, yet the supported workloads operate at higher energy efficiency.
Alternatively, the authors of~\cite{ima_ssc2022} propose an architecture with a scalable array of cores that can trade off energy vs. accuracy. 
They integrate a 4-bit in-memory computing (IMC) accelerator with a near-memory single-instruction multiple-data (SIMD) digital accelerator with flexible accuracy to minimize energy consumption.
Similarly, the authors of~\cite{hp_ssc2020, ueyoshi2022diana} propose SoCs which include a control unit (the CPU) and a series of accelerators that offer a trade-off of accuracy vs. energy efficiency.
In \cite{hp_ssc2020}, the authors include a 590k IMC accelerator specialized in 1-bit operations and a configurable digital near-memory-computing (NMC) accelerator for scalable precision (1-8-bit) computation.

In this work, we use the DIANA SoC~\cite{ueyoshi2022diana} as a benchmarking platform. 
It features a single-core RISC-V host CPU and two accelerators with multiple local memories accessed through Direct Memory Access (DMA). Sec. \ref{sec:methods} gives further details about the DIANA architecture.

\subsection{Software DNN deployment tools}
The diversity in specialized hardware is also reflected in the development of a plethora of different DNN deployment frameworks.
They aim to optimize the execution of a collection of manually or (semi-)automatically optimized routines (called kernels) on a target hardware architecture.

TFLite Micro~\cite{tflitemicro2021} is among the first frameworks introduced for edge AI deployment. 
It allows converting a TensorFlow model to a selected set of hand-optimized C++ NN kernels to run on a minimal C++ runtime for MCUs.
MCUNetv2~\cite{mcunetv2} contains both an optimized runtime and kernel generator that executes layers in a depth-first fashion~\cite{8667835} to reduce peak memory consumption.
Unfortunately, the generality of these frameworks does not allow them to exploit platform-specific optimizations, like memory planning and code generation for embedded accelerators.

DORY~\cite{burrello2021dory} optimizes the memory traffic for DNN deployment on specialized edge devices.
By generating C code that tiles the execution of a dedicated kernel library, DORY reduces the size of intermediate buffers.
This is crucial since microcontrollers often have limited level-1 (L1) memory.
To achieve this, DORY formalizes tiling as an optimized constraint programming problem with kernel-specific heuristics.
The produced code is more optimized but less general than previous solutions. 
Using DORY on a new architecture requires creating a new dedicated kernel library, new templates, and reprogramming the tiler to tailor it to specific hardware.

A popular DNN deployment framework on high-performance and edge devices that alleviates library generation efforts is TVM~\cite{chen2018tvm}. 
TVM's primary optimization mechanism is autotuning: it quickly compiles differently-scheduled yet equivalent kernel implementations, and after running those on hardware, the most optimal kernel is selected.
As such, TVM can implicitly improve the execution time on CPUs and GPUs and fine-grained general matrix multiply (GEMM) accelerators like VTA~\cite{moreau2019hardware}.
Moreover, TVM's runtime can link in (vendor-provided) optimized kernels in LLVM IR, CUDA C, C/C++ into a standalone artifact with the bring your own codegen (BYOC)~\cite{chen2021bring} infrastructure.
However, using TVM's autotuning pipeline is impractical for specialized coarse-grained accelerators since proving coarse-grained kernel equivalence requires complex loop nest analysis.
This can be bypassed by using BYOC, but in this way, many of the automatic optimization opportunities presented by the framework are lost.

To overcome the problem of heterogeneous compilation without creating ``yet another custom toolchain'', we propose \textit{HTVM}. 
HTVM uses a hybrid flow that combines the best of both worlds: it can exploit hardware accelerators by using DORY to perform specialized C code generation and layer tiling, and it leverages TVM's general codegen for creating fused C kernels that perform operations not supported by the accelerator on a regular CPU.
Notably, HTVM operates entirely ahead of time and requires no costly online autotuning. 
\section{Compiler for Heterogeneous Computing}
\label{sec:methods}
\begin{figure}
    \centering
    \includegraphics[scale=0.85]{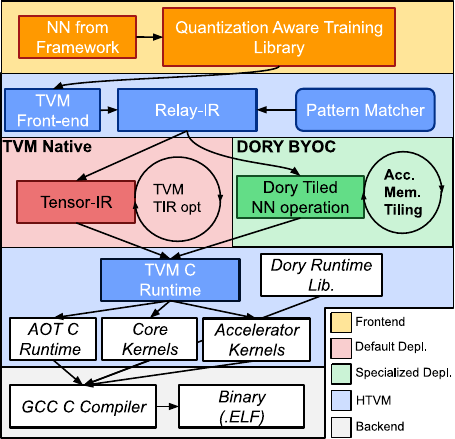}
    \caption{HTVM compilation flow.}
    \label{fig:compilation_flow}
    \vspace{-0.5cm}
\end{figure}

\begin{figure}[t]
    \centering
\includegraphics[width=0.99\columnwidth]{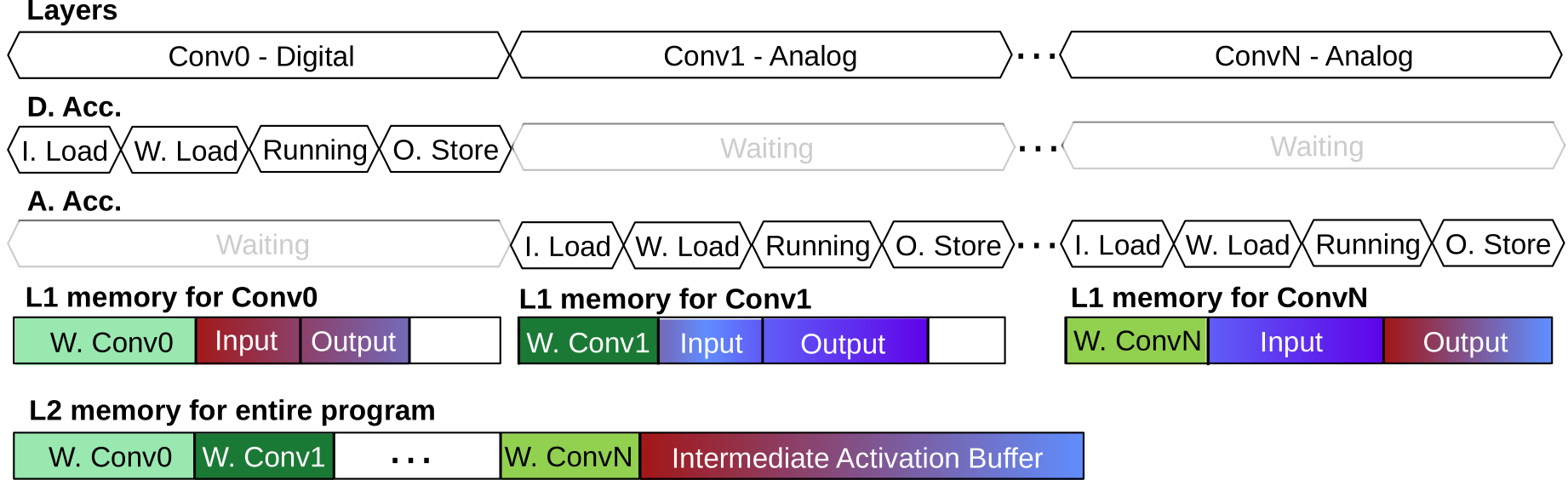}
    \caption{Time diagram of a neural network deployed with HTVM.}
    \label{fig:time_diagram}
    \vspace{-0.7cm}
\end{figure}

HTVM, depicted in Fig. \ref{fig:compilation_flow}, ingests a quantized DNN graph (yellow block) in common formats like TFLite or ONNX with TVM's front end.
The ingested graph is translated into Relay intermediate representation (IR) used by TVM to perform initial optimizations, such as constant folding.
Afterward, an accelerator-aware pattern matcher searches for operator patterns in the Relay graph that the accelerator supports as a single coarse-grained operator, e.g., an 8-bit 2D convolution (Conv2D) followed by a 32-bit bias-add, 8-bit re-quantization, and ReLU. 
Matched patterns are dispatched to the \textit{BYOC DORY} compiler backend (green block), which generates C-code to drive the accelerator. 
Unmatched operators left in the graph follow TVM's native lowering pipeline (red block), which produces operator-fused CPU kernels instead.
Finally, we use TVM to generate a single C function that executes all kernels sequentially. HTVM also yields a memory schedule for allocating and de-allocating intermediate activation tensors in main memory (L2). A timing diagram of a network flow produced by HTVM is reported in Fig. \ref{fig:time_diagram}.

The remainder of this section will focus on two critical enablers for our hybrid flow: (1) a rule-based dispatching mechanism that supports offloading to multiple accelerators with different capabilities, and (2) the DORY backend that manages the accelerator while optimizing the accelerators' local memory (L1) usage.
Deployment with HTVM towards a real heterogeneous SoC, DIANA, concludes the section.

\subsection{Accelerator-aware dispatching}

Our dispatching mechanism is based on a pattern matcher and accelerator-aware rules. The pattern matcher determines possible candidate patterns in the Relay IR graph that can be offloaded to dedicated hardware.
Our implementation uses the TVM BYOC flow \cite{chen2021bring} and the Relay Pattern Matching language. Listing \ref{listing:pattern_matching} provides an example for a \verb+Conv2D-BiasAdd-ReQuant-ReLU+ pattern matcher of the coarse-grained 2D convolution instruction followed by bias addition, re-quantization, and the ReLU activation function.
The accelerator-aware rules describe the constraints of the accelerator in more detail and make the final decision whether a pattern is sent to an accelerator or not, checking if all the parameters (e.g., stride, kernel size, data layout, parameters ranges, and bit-width, etc.) are supported by the accelerator.
If a pattern satisfies all rules of one of the accelerators, the operations will be offloaded to it, and the accelerator-specific flow is employed.
When multiple accelerators on the platform can execute the pattern, the flow selects the one best optimized for that given operation. This choice is based on factors like bit widths, layer geometries, or other user-defined parameters.
When a pattern is not ``matched'', the native TVM flow generates general C code kernels that are executed on the CPU.
Otherwise, the BYOC DORY backend takes on the translation of the selected kernels. 

\subsection{BYOC DORY: Accelerator-aware code generation}
\label{sec:byocdory}

To enable code generation for hardware accelerators while optimizing memory and compute unit utilization, we integrated the open-source DORY \cite{burrello2021dory} framework in the HTVM flow.
DORY's input is a DNN layer that has to be executed on a heterogeneous platform with a general-purpose CPU that drives one/many accelerators and manages a multi-level memory system.
For this input, DORY (1) produces a suitable optimized tiling solution to fit the accelerators' memory (L1), 
(2) generates accelerator-specific and memory-specific  instructions, (3) stores the weights in the SoC's global memory (L2) in the most optimal data layout (i.e., to avoid CPU data-marshaling overheads), and (4) emits an explicit memory management schedule to move the data between different memory levels (L1 and L2).

First, DORY's layer analyzer calls the tiling solver. Tiling is needed whenever a layer does not fit into L1 memory,  which is typically the case in resource-constrained, edge systems \cite{burrello2021dory}.
While maximizing resource (i.e., memory and compute units) utilization (Equation \ref{eq:maxl1}), DORY's tiler ensures that the memory mapping meets the platform's constraints at all times, as expressed in Eq. \ref{eq:constraintl1w}:
\begin{gather}
    \label{eq:maxl1}
    \max(\alpha(L_1^{weight} + L_1^{out} + L_1^{in}) + \sum_i \beta_i \mathcal{H}_i) \\
    \label{eq:constraintl1w}
    L_1^{weight} + L_1^{in} + L_1^{out} < L_1^{A}
\end{gather}
where $L_1^{\{weight,in,out\}}$ indicates the amount of accelerator memory allocated for the weights, inputs, and outputs, $L_1^{A}$ represents the memory of the accelerator, and $\mathcal{H}_i$ are accelerator-aware heuristics, which help to maximize the accelerator utilization. Hyperparameters $\alpha$ and $\beta$ control the balance between maximizing memory utilization and maximizing platform-specific heuristics.
Sec. \ref{sec:diana} details these heuristics for a specific heterogeneous target platform, the DIANA SoC.

Finally, based on the dimensions of the tiles, the layer generator creates code that  performs  weight allocation and memory management and drives the platform's accelerators.
\begin{listing}[tb]
\begin{minted}[mathescape,
              linenos,
              escapeinside=||,
              %highlightlines={4},
              numbersep=5pt,
              gobble=2,
              frame=lines,
              fontsize=\footnotesize,
              framesep=2mm]{python}
conv2d_pattern():
    conv2d = is_op("nn.conv2d")(
        wildcard(), wildcard())
    bias_add = is_op("nn.bias_add")(
        conv2d, wildcard())
    right_shift = is_op("right_shift")(
        bias_add, is_constant())
    clip = is_op("clip")(right_shift)
    cast = is_op("cast")(clip).
    has_attr({"dtype": "int8"})
    act_or_cast = cast.optional(is_op("clip")(x))
    return act_or_cast
\end{minted}
    \vspace{-0.4cm}
    \caption{Pattern Matching code for \texttt{Conv2D-BiasAdd-ReQuant-ReLU}.}
    \label{listing:pattern_matching}
 \end{listing}

\subsection{HTVM flow on DIANA SoC}
\label{sec:diana}
\begin{figure}
    \centering
    \includegraphics[width=\columnwidth]{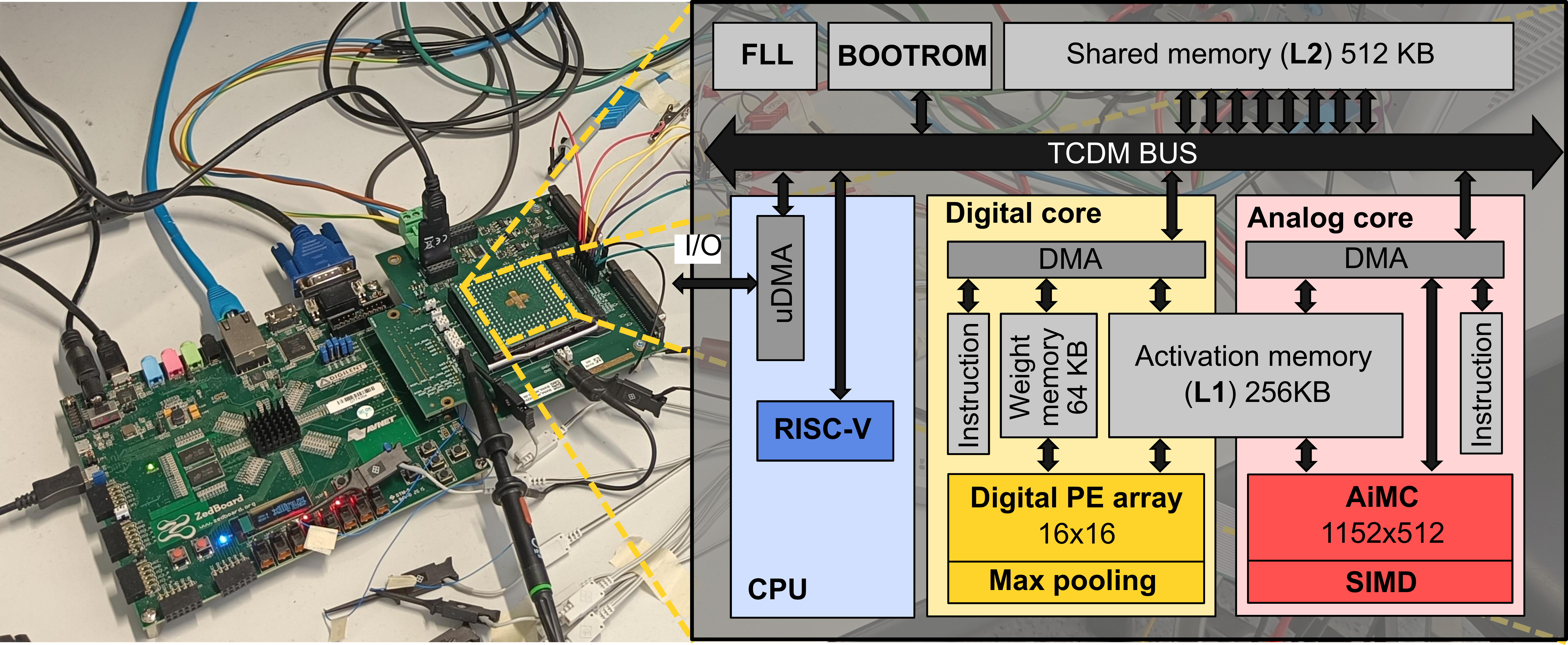}
    \caption{Measurement setup and DIANA architecture from \cite{ueyoshi2022diana}. }
    \label{fig:measurement_setup}
    \vspace{-0.7cm}
\end{figure}
To support a specific heterogeneous platform, the user has to provide to HTVM only three components, i.e., (1) the hardware specifications (memory dimensions, number and type of accelerators, etc.) and operations supported by the dedicated hardware, (2) the heuristics to maximize the accelerator utilization and (3) the platform-specific instructions to call accelerators and manage the memory through DMA calls.

In this section, we briefly describe these three components for DIANA, the platform used to benchmark HTVM, in section \ref{sec:results}.
DIANA is a heterogeneous platform that integrates a RISC-V (RV32IMCFXpulpV2) microcontroller as a primary computational node and two accelerators: the first is based on a 2D SIMD array of 16$\times$16 processing elements (PE) which delivers up to 256 multiply-accumulate (MAC) operations in 8-bit precision per cycle. 
The accelerator also allows executing re-quantization, ReLU, and some pooling operations at the output. 
The Conv2D layers are mapped unrolling the output channels ($K$) and output feature map width ($o_x$) in the two physical dimensions of the array; fully-connected (FC) layers are instead unrolled along input channels ($C$) and $K$.
The second accelerator is an analog IMC accelerator, which embeds an array of 1152$\times$512 SRAM cells to execute MAC operations with 7-bit inputs and ternary weights. Also, this accelerator supports batch normalization, residual addition, pooling, activation functions, and re-quantization.
Both accelerators share a 256kB L1 input/output memory, and each has its weight memory (64kB for digital and 144kB analog).

To support DIANA in HTVM, we add (1) memory sizes, specific instruction formats, etc. to DORY and supported operators to the Relay pattern matcher.
Since both accelerators support convolutions, we discern which accelerator to use by simply looking at the provided weights' bit-width of the convolution:
8-bit precision goes to digital, and ternary precision goes to analog.
For (2), we design a series of heuristics to support the optimization of individual accelerator utilization and memory transfers.
To maximize analog accelerator utilization, we spatially unroll $C$ and $K$ as much as possible.
For the digital accelerator, we favor the $C$ and $i_x$ (input width) dimensions to be multiples of 16 to exploit all the PEs (16 rows and 16 columns).
To minimize non-contiguous input data transfers on the digital accelerator, since they are stored and processed  using $C-y-x$ layout, we maximize the $i_y$ (input height) dimension.
As an example, the heuristics that DORY maximizes to optimize DIANA's digital accelerator utilization are:
\begin{align}
   &\label{eq:h_pe_digital_K}\mathcal{H}_{pe\_digital\_C} = (C^t - 1)\mod{16} \\
   &\label{eq:h_pe_digital_Oy}\mathcal{H}_{pe\_digital\_ix} = (i^t_x - 1)\mod{16}\\
   &\label{eq:h_dma}\mathcal{H}_{DMA} = i^t_y.
\end{align}
Where $C^t$ is the tile size for the input channel dimensions of the computation, and $i^t_y$ and $i^t_x$ are the input width and height dimension tile size, respectively.
Each optimization objective $\mathcal{H}_i$ is then combined in eq. \ref{eq:maxl1}. 
Note that even though the accelerators can execute arbitrarily sized input feature maps, the tile sizes strongly influence the overall accelerator spatial utilization, as shown in Sec. \ref{sec:results}.

Finally, for the DIANA-specific instructions (3), we implemented dedicated libraries to offload computation to the analog and digital accelerators.
\section{Experimental Results}
\label{sec:results}

Throughout this section, we show the flexibility and performance of our proposed compilation flow, optimizations, and lightweight runtime.
As mentioned earlier, DIANA is used as the target platform.
All the benchmarks were directly deployed on the DIANA platform (see our measurement setup in Fig. \ref{fig:measurement_setup}).
Cycles and latencies are measured at 260MHz with dedicated hardware performance counters on the RISC-V core.
This section starts by demonstrating the possibilities of hardware-aware tiling.
Afterward, we benchmark single-layer execution on both accelerators.
The final subsection highlights HTVM's end-to-end performance on the MLPerf\textsuperscript{TM} Tiny benchmarks exploiting both accelerators and compares this with the SotA.

\subsection{Hardware-aware Tiling}
\begin{figure}
    \centering
    \includegraphics[width=\columnwidth]{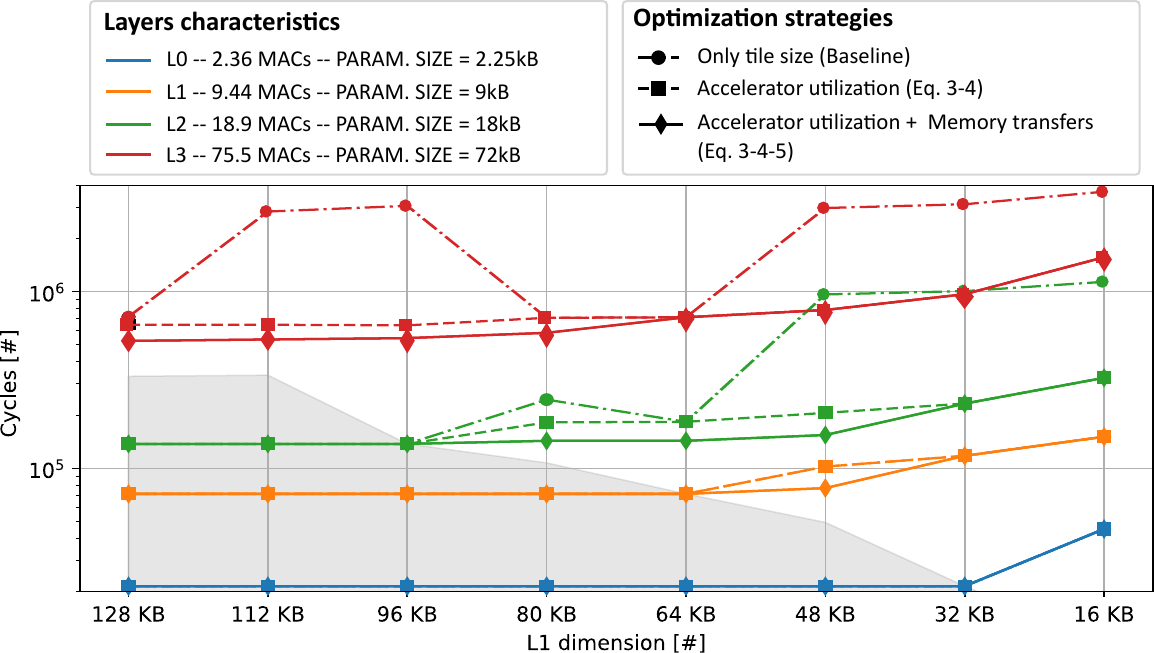}
    \caption{Latency effect of tiling with accelerator-aware heuristics for \textit{decreasing} L1 memory budget to execute different layers on DIANA's digital accelerator. }
    \label{fig:tiling}
    \vspace{-0.7cm}
\end{figure}
The first important feature of HTVM is its ability to leverage DORY's accelerator-aware tiling to fit the execution of large layers into TinyML systems with small memories.
Fig. \ref{fig:tiling} shows the latency required to execute convolutional layers, of different sizes with increasing memory constraints, adopting different optimization approaches for tiling.
The different markers in the figure indicate the number of cycles for memory tiling with different heuristics. The round markers indicate the cycle count of tiled layers without any applied heuristics; square and diamond ones additionally use the heuristics Eq. \ref{eq:h_pe_digital_K}, \ref{eq:h_pe_digital_Oy} and Eq. \ref{eq:h_pe_digital_K}, \ref{eq:h_pe_digital_Oy}, \ref{eq:h_dma} respectively.
Note that solutions whose evaluation falls in the grey area do not require tiling, as the local memory is big enough to host the entire size of the layer.
The figure clearly shows that applying both heuristics incurs lower or equivalent cycle counts in all experiments.
From the round markers, we can notice that, when not adding any heuristics to the optimizer, we could either have good tiles (e.g., in the middle of the red curve) or very bad tiles (beginning of the red curve), given that the utilization of the accelerator could be either high or low, as it is not taken into account during optimization.
The advantage of employing the heuristics is indeed evident in all curves, reaching up to 6.2$\times$ faster execution.

\subsection{Single Layer Overhead Characterization}
\label{sec:overhead}
\begin{figure}
    \centering
    \includegraphics[width=\columnwidth]{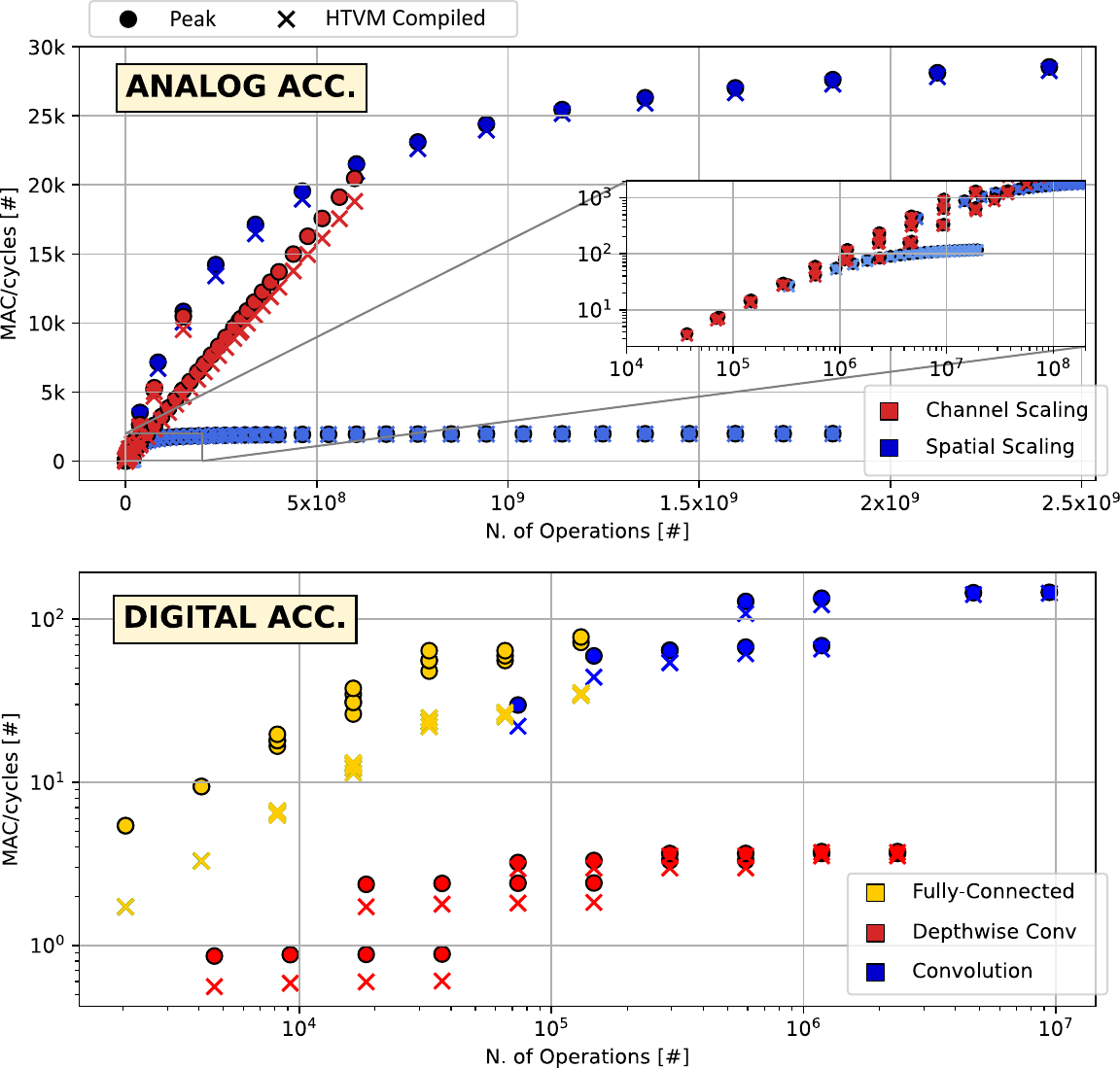}
    \caption{Single layer overhead characterization on digital and analog accelerators with Conv2D, FC, and DWConv2D layer types, evaluated for different geometries.
    For the analog layers, a distinction is made between scaling the channels, or the spatial dimension to explore different geometries. For the digital layers, we explore spatial scaling with Conv2D, and channel scaling with FC layers.}
    \label{fig:single_layer}
    \vspace{-0.7cm}
\end{figure}

Performing a DNN layer on a DIANA accelerator requires some setting up, handshaking, and shutting down/cleaning up.
For example, in the case of a tiled convolution, a loop iterates over all tiles, and each individual tiled call has to move data from the main memory (L2) to the accelerator input/output memory (L1).
To characterize this overhead, we profiled generated kernels of various kinds and geometries on the two accelerators in two ways.
On one hand, we have the peak performance of the accelerator, i.e., the measured time between the triggering of the accelerator and its completion.
On the other hand, we have the full kernel call as generated by HTVM, measured between the call and return on the RISC-V host. 
Note that the weights transfer is also included in the peak performance, as this transfer is orchestrated in the same instruction to execute the layer.

Fig. \ref{fig:single_layer} (top) shows different geometries by increasing the channels or the spatial dimension of the layer (at different no. of MACs of convolutional (Conv2D) layers and the equivalent execution throughput on the analog core.
The lower plot of Fig \ref{fig:single_layer} shows the same setup for the digital core, which also supports (FC) and depthwise convolutional (DWConv2D) layers.
For Conv2D layers, both graphs show a small overhead.
On the analog core, the average throughput loss is about 5.20\% on average, with a minimum of 0.51\% for highly computationally intensive layers.
The digital Conv2D loses at best only 1.32\% in throughput. 
On the other hand, the throughput loss for the fastest FC layer is about 54.5\%. 
The reason is two-folded: for small layers, given the very low amount of computation cycles of the digital accelerator, the overhead of the HTVM runtime is more tedious. Bigger layers require many tiling iterations (given the low arithmetic intensity), which results in overhead for DMA calls for input/output buffers. 
DWConv2D layers use only one row of PEs on DIANA's digital accelerator at a maximum peak throughput of 3.75MACs/cycle.
Here the full kernel is never more than 20.7\% slower. 
Overall, we can notice that for more common and expensive workloads (i.e., Conv2D), we keep a very low overhead, meaning that HTVM can reach excellent performance on end-to-end networks, as shown in the next section.

\subsection{Networks: TinyML Benchmarks}

To characterize the end-to-end performance of our approach, we deployed the four networks of the MLPerf\textsuperscript{TM} Tiny Benchmark suite version 1.0\cite{mlcommons}. 
This suite includes four topologies: a ToyAdmos Deep Auto Encoder (DAE), an audio processing CNN (DS-CNN), MobileNet V1, and a CIFAR10 ResNet image classifier. 
These models represent typical edge platform workloads, including (DW)Conv2D, FC, element-wise addition, average pooling, and softmax layers.

\begin{table}[t]
\caption{Latency and binary size of deployed MLPerf\textsuperscript{TM} Tiny benchmarks on the DIANA SoC in different configurations.}
\label{tab:networkperformance}
\begin{tabular}{|l|r|p{0.4cm}p{0.4cm}|p{0.5cm}p{0.7cm}|p{0.5cm}p{0.7cm}|}
\hline
Platform            & \multicolumn{7}{|c|}{DIANA SoC}                                                                       \\ \hline
Core    & CPU &\multicolumn{2}{|c|}{CPU + Dig.}           & \multicolumn{2}{c|}{CPU + Ana.}  & \multicolumn{2}{c|}{CPU + Both}   \\\hline
Compiler& \multicolumn{1}{|c|}{TVM}     & \textit{Peak}       & HTVM      & \textit{Peak}    & HTVM          & \textit{Peak}    & HTVM                                  \\\hline
 \multicolumn{8}{l}{\textbf{DSCNN - Keyword Spotting}\dag}                                                     \\\hline
Lat. (ms)        &  48.24 & \textit{1.70}    & 1.75     &     \textit{13.51}      & 13.51      &      \textit{1.66}          & \textbf{1.69}\\
Size (kB)    &  \textbf{59}      & \textit{59}          & 60     &     \textit{93}          & 93          &   \textit{81}           & 81             \\ \hline
 \multicolumn{8}{l}{\textbf{MobileNet - Visual Wake Words}}                                                    \\\hline
Lat. (ms)        & OoM*       & \textit{5.42}       &  \textbf{5.68}    &    \textit{40.67}         & 40.67          &      \textit{5.39}         & 5.82\\
Size (kB)    & 289       & \textit{305}         &  306       &      \textit{239}         & \textbf{239}             &      \textit{292}         & 293    \\ \hline
 \multicolumn{8}{l}{\textbf{ResNet - Image Classification}}                                                 \\\hline
Lat. (ms)        & 134.11   & \textit{0.66}         &  1.19     &       \textit{1.52}        & 1.53         &       \textit{0.61}        & \textbf{1.12}\\
Size (kB)    & 122     & \textit{107}         &  \textbf{107}     &  \textit{129}             & 129         &       \textit{107}        & 108   \\ \hline
 \multicolumn{8}{l}{\textbf{ToyAdmos - Anomaly Detection} }                                                   \\\hline
Lat. (ms)        & 4.70     & \textit{0.30}         &  \textbf{0.36}     &   \textit{0.80}             & 0.80         &    \textit{0.49}           & 0.52\\
Size (kB)    & 287     & \textit{315}         &  315     &     \textit{171}          & \textbf{171}         &     \textit{274}          & 275            \\ \hline
\multicolumn{8}{l}{*Out of Memory (OoM) \dag Input filter size adapted to [7,5]}
\end{tabular}
    \vspace{-0.7cm}
\end{table}

HTVM's flexibility allows quickly deploying each of these networks to different configurations of DIANA, exploiting one or both of the accelerators.
The columns of table \ref{tab:networkperformance} show binary size and latency for each deployment scenario discussed below.
All C code is \texttt{-O3}-compiled with an XpulpV2-aware RISC-V GCC.
Peak measurements are done in the same fashion as in section \ref{sec:overhead} and do not affect TVM-generated kernels, to be 
indicative of how effective HTVM can utilize the accelerators.
As a performance baseline on DIANA, the Digital TVM (8-bit) configuration only uses DIANA's RISC-V core.
In this configuration, MobileNet stops running with an error, since more than 512kB of memory has to be allocated.

In the Digital HTVM configuration, all (DW)Conv2D, FC, and Add layers are offloaded to DIANA's 8-bit digital accelerator.
For ResNet, this results in an impressive speedup of 112$\times$ over the TVM baseline. 
At the same time, the binaries shrink for 2 benchmarks since DIANA's coarse-grained accelerator requires fewer instructions than the RISC-V core to perform certain operators.
This effect is most pronounced in ResNet, where a reduction of 12.3\% is achieved at equal bit-precision. 

In the ``Analog'' column, we display the performance of ternary layers on DIANA's analog accelerator, which only supports Conv2D layers. 
We can still deploy all networks by implementing FC layers as Conv2Ds and by offloading the other unsupported layers with TVM to the RISC-V core in 8-bit\footnote{TVM does not support generating ternary kernels.} albeit at a much lower performance.
MobileNet and DS-CNN suffer from this performance drop because of their DWConv2D layers.
Also, on the other two networks, we still have worse performance compared to the digital accelerator, given the low amount of channels of these networks and the overhead of filling the analog accelerator weight memory for each layer.
As can be seen for ToyAdmos and MobileNet, ternary weights data require less storage.
In the other networks on DIANA, however, some layer dimensions require padding the L2 memory with zeros to fill a part of the large IMC macro which ultimately leads to a larger overall binary.

HTVM also allows for a ``mixed configuration'' where we combine the abilities of both accelerators. 
The first and last accelerator-eligible layers and all DWConv2D layers are executed digitally, remaining Conv2D's are executed on the analog core. 
The rationale is to execute all the layers that do not cause an accuracy drop on the analog accelerators.
By offloading more kernels to the different accelerators, we achieve the best latency in 2 out of 4 benchmarks.
For instance, the mixed DS-CNN configuration is 8$\times$ faster than its analog counterpart.
Furthermore, the HTVM Mixed ResNet has 120$\times$ improvement over plain TVM deployment.
Overall, we show that HTVM is flexible enough to offload single kernels to different hardware blocks, enabling the easy porting of neural networks with a wide range of different layers on a real state-of-the-art heterogeneous platform.

\subsection{Comparison with state-of-the-art}

\begin{table}[t]
\caption{Performance comparison of deployed MLPerf\textsuperscript{TM} Tiny benchmarks with HTVM and SotA tools and platforms\cite{mlcommons}}
\label{tab:sota}
\begin{tabular}{|l|r|r|r|r|}
\hline
\multirow{2}{*}{Compiler}            & TVM          & \multicolumn{1}{|c|}{TVM +}          & GreenWaves         & \multicolumn{1}{|c|}{\textbf{HTVM}}                        \\
                    &              & CMSIS-NN   & \multicolumn{1}{|c|}{GAPFlow}            & (\textbf{This paper})                 \\\hline
Platform**           & \multicolumn{2}{|c|}{STM32L4R5ZIT6U}      & \multicolumn{1}{|c|}{GAP9}    & \multicolumn{1}{|c|}{DIANA Digital}   \\ \hline
 \multicolumn{5}{l}{\textbf{Benchmarks - Latency* (ms)}}                                                                             \\ \hline
DSCNN               &   66.6              &        46.1               &         0.68                      &   1.75                                \\ \hline
MobileNet           &   155                &       139                &         1.61                      &   5.68                                \\ \hline
ResNet            &   180                &       180                &         0.88                      &   1.19                                \\ \hline
ToyAdmos            &    5.4               &        3.97               &        0.256                       &   0.36                                \\ \hline
\multicolumn{5}{l}{*Normalized for the same clock frequency (260 MHz)} \\
\multicolumn{5}{l}{**Fastest HW-SW operating configuration} \\
\end{tabular}
    \vspace{-0.7cm}
\end{table}
Table \ref{tab:sota} compares 8-bit solutions from our work with results submitted at \cite{mlcommons} at a normalized clock frequency of 260MHz.
HTVM on DIANA beats ResNet on a regular STM microcontroller with TVM-generated kernels by 150x lower latency. 
If the same MCU also offloads to CMSIS-NN kernels (an open-source backend for ARM CPUs), HTVM on DIANA can still perform MobileNet inference 24$\times$ faster.
Networks compiled with GapFlow on GAP9, a manually-tuned commercial closed-source TinyML platform, still outperform our results.
Note that while our results are strongly worse on networks containing DW layers (for which DIANA is not optimized), we are closer for the other two benchmarks (35.5\% slower on ResNet).
However, HTVM already performs very competitively without any manual tuning and does not prevent further tuning. 
To enable further tuning on a platform, HTVM can  easily be expanded with other BYOC codegens to deploy hand-tuned CPU kernels. 
\section{Conclusions}
\label{sec:conclusion}
In this paper, we present HTVM, an open-source compiler that combines the strengths of TVM and DORY to allow smooth DNN deployment towards a broad class of accelerators while enabling effective hardware utilization on resource-constrained TinyML platforms.
We show that our fully ahead-of-time flow results in a smaller binary size, orders of magnitude faster execution, and close-to-peak accelerator performance on the DIANA SoC.
We prove this by showing performance competitive with the state of the art on the MLPerf\textsuperscript{TM} Tiny benchmark suite.


\footnotesize
\bibliographystyle{IEEEtran}
\bibliography{bstctl, references}

\begin{thebibliography}{10}
\providecommand{\url}[1]{#1}
\csname url@samestyle\endcsname
\providecommand{\newblock}{\relax}
\providecommand{\bibinfo}[2]{#2}
\providecommand{\BIBentrySTDinterwordspacing}{\spaceskip=0pt\relax}
\providecommand{\BIBentryALTinterwordstretchfactor}{4}
\providecommand{\BIBentryALTinterwordspacing}{\spaceskip=\fontdimen2\font plus
\BIBentryALTinterwordstretchfactor\fontdimen3\font minus
  \fontdimen4\font\relax}
\providecommand{\BIBforeignlanguage}[2]{{%
\expandafter\ifx\csname l@#1\endcsname\relax
\typeout{** WARNING: IEEEtran.bst: No hyphenation pattern has been}%
\typeout{** loaded for the language `#1'. Using the pattern for}%
\typeout{** the default language instead.}%
\else
\language=\csname l@#1\endcsname
\fi
#2}}
\providecommand{\BIBdecl}{\relax}
\BIBdecl

\bibitem{genc2021gemmini}
H.~Genc \emph{et~al.}, ``{Gemmini: Enabling systematic deep-learning
  architecture evaluation via full-stack integration},'' in \emph{DAC}, 2021.

\bibitem{ueyoshi2022diana}
K.~Ueyoshi \emph{et~al.}, ``{DIANA: An End-to-End Energy-Efficient Digital and
  ANAlog Hybrid Neural Network SoC},'' in \emph{ISSCC}, 2022.

\bibitem{jain2022tinyvers}
V.~Jain \emph{et~al.}, ``{TinyVers: A 0.8-17 TOPS/W, 1.7 $\mu$W-20 mW, Tiny
  Versatile System-on-chip with State-Retentive eMRAM for Machine Learning
  Inference at the Extreme Edge},'' in \emph{VLSI}, 2022.

\bibitem{accelerator_comparison}
\BIBentryALTinterwordspacing
G.~Kaiyuan \emph{et~al.} {Neural Network Accelerator Comparison}. [Online].
  Available:
  \url{https://nicsefc.ee.tsinghua.edu.cn/projects/neural-network-accelerator/}
\BIBentrySTDinterwordspacing

\bibitem{burrello2021dory}
A.~Burrello \emph{et~al.}, ``{DORY: Automatic end-to-end deployment of
  real-world DNNs on low-cost IoT MCUs},'' \emph{IEEE Trans Comput.}, 2021.

\bibitem{chen2018tvm}
T.~Chen \emph{et~al.}, ``{TVM: An automated End-to-End optimizing compiler for
  deep learning},'' in \emph{OSDI}, 2018.

\bibitem{xavier_dac22}
I.~Dagli \emph{et~al.}, ``Axonn: Energy-aware execution of neural network
  inference on multi-accelerator heterogeneous socs,'' in \emph{DAC}, 2022.

\bibitem{hp_ssc2020}
H.~Jia \emph{et~al.}, ``A programmable heterogeneous microprocessor based on
  bit-scalable in-memory computing,'' \emph{JSSC}, 2020.

\bibitem{ima_ssc2022}
H.~Jia \emph{et~al.}, ``Scalable and programmable neural network inference
  accelerator based on in-memory computing,'' \emph{JSSC}, 2022.

\bibitem{tflitemicro2021}
R.~David \emph{et~al.}, ``{TensorFlow Lite Micro: Embedded Machine Learning for
  TinyML Systems},'' in \emph{MLSys}, 2021.

\bibitem{mcunetv2}
J.~Lin \emph{et~al.}, ``Mcunetv2: Memory-efficient patch-based inference for
  tiny deep learning,'' 2021.

\bibitem{8667835}
K.~Goetschalckx \emph{et~al.}, ``{Breaking High-Resolution CNN Bandwidth
  Barriers With Enhanced Depth-First Execution},'' \emph{JETCAS}, 2019.

\bibitem{moreau2019hardware}
T.~Moreau \emph{et~al.}, ``A hardware--software blueprint for flexible deep
  learning specialization,'' \emph{Micro}, 2019.

\bibitem{chen2021bring}
Z.~Chen \emph{et~al.}, ``{Bring Your Own Codegen to Deep Learning Compiler},''
  \emph{arXiv preprint arXiv:2105.03215}, 2021.

\bibitem{mlcommons}
\BIBentryALTinterwordspacing
{MLPerf Tiny Benchmark V1.0 Results}. [Online]. Available:
  \url{https://mlcommons.org/en/inference-tiny-10/}
\BIBentrySTDinterwordspacing

\end{thebibliography}

\end{document}